\documentstyle[12pt,epsf]{article}
\newcommand{\lsm}{L$\sigma$M}
\newcommand{\lsms}{\mbox{\scriptsize L$\sigma$M}}

\newcommand{\cls}{\mbox{\scriptsize CL}}
\newcommand{\cms}{\mbox{\scriptsize CM}}
\newcommand{\qts}{\mbox{\scriptsize QT}}
\newcommand{\slw}{K^+\rightarrow\pi^0e^+\nu}
\newcommand{\refc}[1]{Ref.~\cite{#1}}
\newcommand{\ha}{\frac{1}{2}}

\newcommand{\dbarfp}{\,\bar{}\:\!\!\!\!d\;\!^4\:\!\!p}
\newcommand{\abs}[1]{\left| #1\right|}
\newcommand{\av}[1]{\left<#1\right>}
\newcommand{\bk}[2]{\left<#1|#2\right>}
\newcommand{\bko}[3]{\left<#1|#2|#3\right>}
\newcommand{\bm}[1]{\mbox{\boldmath $#1$}}
\newcommand{\be}{\begin{equation}}
\newcommand{\ee}{\end{equation}}
\newcommand{\bea}{\begin{eqnarray}}
\newcommand{\eea}{\end{eqnarray}}
\newcommand{\eqr}[1]{Eq.~(\ref{#1})}
\newcommand{\eqrs}[2]{Eqs.~(\ref{#1}) and (\ref{#2})}
\newcommand{\fnd}[2]{\frac{\textstyle #1}{\textstyle #2}}

\newcommand{\fsigma}{\mbox{$f_{0}$(600)}}
\newcommand{\fn}{\mbox{$f_{0}$(980)}}
\newcommand{\ft}{\mbox{$f_{0}$(1370)}}
\newcommand{\ff}{\mbox{$f_{0}$(1500)}}

\begin{document}\baselineskip .7cm
\title{\bf Meson Form Factors and \\ the Quark-Level Linear $\sigma$ Model}
\author{
Michael D.\ Scadron$^{\,a}\!\!$\,,
Frieder Kleefeld$^{\,b}\!\!$\,, 
George Rupp$^{\,b}\!\!$\,, and
Eef van Beveren$^{\,c}\!\!$ \\[5mm]
$^{a}${\footnotesize\it Physics Department, University of Arizona, Tucson,
AZ 85721, USA} \\ {\footnotesize\tt scadron@physics.arizona.edu} \\[.3cm]
$^{b}${\footnotesize\it Centro de F\'{\i}sica das Interac\c{c}\~{o}es
Fundamentais, Instituto Superior T\'{e}cnico, P-1049-001 Lisboa, Portugal} \\
{\footnotesize{\tt kleefeld@cfif.ist.utl.pt}, 
{\tt george@ajax.ist.utl.pt} (corresponding author)} \\[.3cm]
$^{c}${\footnotesize\it Departamento de F\'{\i}sica, Universidade de Coimbra,
P-3004-516 Coimbra, Portugal} \\ {\footnotesize\tt eef@teor.fis.uc.pt} \\[5mm]
[.3cm]
{\small hep-ph/0211275}
}
\date{\today}
\maketitle

\begin{abstract}
The quark-level linear $\sigma$ model (\lsm) is employed to compute a variety
of electromagnetic and weak observables of light mesons, including pion and
kaon form factors and charge radii, charged-pion polarizabilities, semileptonic
weak $K_{\ell3}$ decay, semileptonic weak radiative pion and kaon form factors,
radiative decays of vector mesons, and nonleptonic weak $K_{2\pi}$ decay.
The agreement of all these predicted observables with experiment is striking. In
passing, the tight link between the \lsm\ and vector-meson dominance is
shown. Some conclusions are drawn on the \lsm\ in connection with lattice and
renormalization-group approaches to QCD.
\end{abstract}
\section{Survey of L\bm{\sigma}M and chiral Goldberger--Treiman relations}
For the past eight years, there has been much experimental \cite{PDG02} and
theoretical \cite{TH95_99} activity, as well as combined workshops
\cite{WS00_02},
concerning isoscalar scalar mesons in general, and the $\sigma$ meson in
particular. Very recently, we have employed electromagnetic (e.m.) and weak
processes to conclude that the mostly nonstrange $\bar{n}n$ resonances are the
\fsigma\ and the \ft, while the \fn\ and the \ff\ are mainly $\bar{s}s$
\cite{KBRS02}. In the present paper, we study meson ($\pi$,$K$) form factors
in general, and specialize at a later stage to a specific scheme, namely the
quark-level linear $\sigma$ model (\lsm).

Nonperturbatively solving \cite{DS95} the strong-interaction Nambu-type gap
equations
$\delta f_\pi=f_\pi$ and $\delta\hat{m}=m$ (where $f_\pi$ is the pion decay
constant and $\hat{m}$ is the nonstrange constituent quark mass) in quark-loop
order, regularization schemes lead to the NJL \cite{NJL61} and $Z=0$
compositeness \cite{SWS62_98} relations
\be
m_\sigma \; = \; 2\hat{m} \;\;\; , \;\;\; g \; = \; \frac{2\pi}{\sqrt{N_c}}\;,
\label{comp}
\ee
with $\hat{m}\sim M_N/3$ and meson-quark coupling $g=2\pi/\sqrt{3}=3.628$. For
a more detailed description of the quark-level \lsm, see the Appendix. Here, we
survey instead meson form factors and related data in a \lsm\ context for
strong, e.m., and weak interactions.

This chiral \lsm\ is based on the quark-level pion and kaon Goldberger--Treiman
relations (GTRs)
\begin{equation}
f_{\pi}\,g \; = \; \hat{m} \;= \; \ha(m_u+m_d) \;\;\; , \;\;\; f_K\,g \; = \;
\ha(m_s+\hat{m}) \; ,
\label{gtrs}
\end{equation}
for $f_{\pi}\approx93$ MeV ($f_\pi\approx90$ MeV in the chiral limit (CL) 
\cite{CS81}), $f_K/f_{\pi}\approx1.22$, and $m_s\approx1.44\,\hat{m}$ (from
\eqr{gtrs}). We begin in Sec.~2 by studying meson vector form factors and
their measured charge radii. In
Sec.~3 we survey charged-pion polarizabilities for
$\gamma\gamma\rightarrow\pi\pi$, and compare the results with the \lsm\
predictions. In
Sec.~4 we study the semileptonic weak $K_{l3}$ decays and the form factor
$f_+(k^2)$ evaluated at $k^2=0$. Then in Sec.~5 we examine the radiative
semileptonic weak form factors for $\pi^+\rightarrow e^+\nu\gamma$ and
$K^+\rightarrow e^+\nu\gamma$ decays, with the observed pion second axial-vector
form factor implying a pion charge radius $r_{\pi}\sim0.6$ fm, also found in
Sec.~2 from data \cite{D82} and from the theoretical \lsm. In Sec.~6 we return
to the \lsm\ and its link with vector-meson dominance (VMD). Finally, in Sec.~7
we begin by studying the $\Delta I\!=\!1/2$ rule for two-pion decays of the
kaon in connection with the $\sigma$ as the pion's chiral partner, and end
by showing that the mass of the now experimentally confirmed scalar $\kappa$
meson is consistent with the observed $K\rightarrow2\pi$ decay rate. We
summarize our results and draw our conclusions in Sec.~8.

\section{Meson vector form factors and charge radii}
The charged-pion and kaon e.m.\ vector currents are defined as
\begin{eqnarray}
\bko{\pi^+(q')}{V^{\mu}_{em}(0)}{\pi^+(q)}
 \; & = & F_\pi (k^2) \; (q' + q)^{\mu} \; , \nonumber \\
 \bko{K^+(q')}{V^{\mu}_{em}(0)}{K^+(q)} & = & F_K (k^2) \;
 (q' + q)^{\mu} \; ,
\label{pikcur}
\end{eqnarray}
with $k_\mu = q'_\mu - q_\mu$. The former pion form factor $F_\pi (k^2)$
can be --- perturbatively --- characterized by the (constituent) quark $udu$
and $dud$ loop graphs of Fig.~1a, while the charged-kaon form factor
\begin{figure}[ht]
\unitlength1cm
\epsfxsize=  11.5cm
\epsfysize=  9cm
\centerline{\epsffile{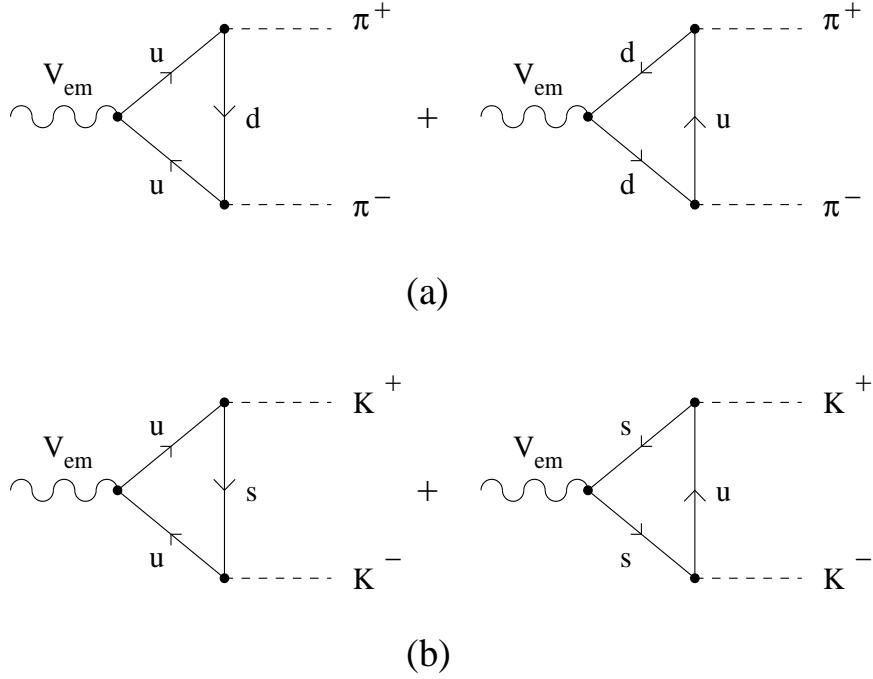}}
\caption{VPP quark triangle graphs.} \label{figgampipi3}
\end{figure}
$F_K (k^2)$ is in a similar manner determined by the $usu$ and $sus$ loop
graphs depicted in  Fig.~1b. Even if each of the diagrams in Fig.~1
appears to be linearly divergent by naive power counting, gauge invariance
enforces every single quark triangle (QT) to be merely logarithmically
divergent. 
After evaluation of spin traces, the form factors in \eqr{pikcur} can be
brought to the form (with color number $N_c = 3$)
\begin{eqnarray}
\label{fpiqt}
F_\pi (k^2)_{\qts} & = & - \, 4ig^2  N_c \, \left( +
\,\frac{2}{3} \, I(k^2,m^2_u,m^2_d,m_\pi^2)
+ \,\frac{1}{3} \, I(k^2,m^2_d,m^2_u,m_\pi^2) \right) \; ,  \\
F_K (k^2)_{\qts} & = & - 4ig^2  N_c \, \left( +
 \,\frac{2}{3} \, I(k^2,m^2_u,m^2_s,m_K^2)
+ \,\frac{1}{3} \, I(k^2,m^2_s,m^2_u,m_K^2) \right) \; .
\label{fkqt}
\end{eqnarray}
The integral $I(k^2,m^2_q,m^2_{{}_Q},M^2)$ is defined by
\begin{eqnarray} \lefteqn{I(k^2,m^2_q,m^2_{{}_Q},M^2) \quad =} \nonumber \\
 & = & \frac{i \pi^2}{(2\pi)^4} \, \frac{1}{2} \int^1_0 dv \int^1_v du \,
 \frac{k^2 u + 2 \, ( M^2 -  (m_q - m_{{}_Q})^2 )  (1-u)}{m^2_Q -
 \Big( M^2 + m^2_Q - m^2_q\Big)  u  + M^2  u^2  + \frac{1}{4} \,
(v^2 - u^2) k^2} \nonumber \\ & & \nonumber \\
 & & + \; \int^1_0 dx \; \int d\!\!{}^{- 4} p\; \Big[ p^2 - m^2_Q  +
 \Big( M^2 + m^2_Q - m^2_q\Big) x  - M^2  x^2 
 \Big]^{-2} \; ,
\label{ikmqq}
 \end{eqnarray}
where $d\!\!{}^{- 4} p = d^4p \; (2\pi)^{-4}$.

The perturbative QT expressions (\ref{fpiqt})--(\ref{ikmqq})
in the CL (i.e. $M\rightarrow 0$) should be compared to
a CL non-perturbative L$\sigma$M result \cite{DS95,PS83}
\begin{eqnarray}
\label{fpilsm}
F_\pi (k^2)^{\cls}_{\lsms} & = & -  4 i 
g^2  N_c \, \int^1_0 dx \int d\!\!{}^{- 4} p\; \Big[ p^2 -  \hat{m}^2  +
 x (1-x)  k^2 \Big]^{- 2} \; ,  \\
F_K (k^2)^{\cls}_{\lsms} & = & -  4 i 
 g^2  N_c \,\int^1_0 dx \int d\!\!{}^{- 4} p\; \Big[ p^2 -  m_{us}^2  +
 x (1-x)  k^2  \Big]^{- 2} \; ,
\label{fklsm}
\end{eqnarray}
where $m_{us} = (m_s + \hat{m})/2$. The logarithmic divergence of these
expressions has been guaranteed through a rerouting procedure \cite{PS83,HS91}.
When $k^2=0$, these form factors become automatically normalized to unity,
i.e.,
\begin{eqnarray}
\label{fpilsm0}
F_\pi (0)^{\cls}_{\lsms} & = & - 4 i
 g^2  N_c \, \int d\!\!{}^{- 4} p\; \Big[ p^2 -  \hat{m}^2 
\Big]^{- 2} \; = \; 1 \; ,  \\
F_K (0)^{\cls}_{\lsms} & = & - 4 i 
g^2  N_c \, \int d\!\!{}^{- 4} p\; \Big[ p^2 -  m_{us}^2 
\Big]^{- 2} \; = \; 1 \; ,
\label{fklsm0}
\end{eqnarray}
due to the GTRs in \eqr{gtrs}, and the definition of the pion and kaon decay
constants $\bko{0}{A^\mu_3}{\pi^0} = i f_\pi q^\mu$,
$\bko{0}{A^\mu_{4-i5}}{K^+} = i\sqrt{2} f_K q^\mu$, with $f_\pi \approx
93$ MeV and $f_K/f_\pi \approx 1.22$ \cite{DS95,HS91}.

In contrast, the perturbative QT results yield in the CL
\begin{eqnarray}
\label{fpiqt0}
F_{\pi^+}(0)^{\cls}_{\qts} & = &  - 4 i  g^2
N_c \, \int d\!\!{}^{- 4} p \; \Big[ p^2 -  \hat{m}^2 
\Big]^{-2} \; ,  \\ & & \nonumber \\
F_{K^+} (0)^{\cls}_{\qts} & = & - 4 i g^2 
N_c \, \Bigg\{ \int d\!\!{}^{- 4} p \; \Big[\Big(p^2  - \hat{m}^2 
\Big) \, \Big(p^2 -  m^2_s\Big)\Big]^{-1}
 \nonumber \\ & & \nonumber \\
 & & \qquad\qquad\quad -\,  \frac{i\pi^2}{(2\pi)^4} \; \frac{1}{2\,
 (m_s+\hat{m})^2} \; \Bigg( m^2_s + \hat{m}^2   
 - \frac{2 m^2_s  \hat{m}^2}{m^2_s - \hat{m}^2} \;\ln \left|
 \frac{m^2_s}{\hat{m}^2} \right|
\Bigg) \Bigg\} \; , \nonumber \\
\label{fkqt0}
\end{eqnarray}
being --- up to an finite constant correction term in the case of the kaon ---
normalized by the logarithmically divergent gap equations (LDGEs)
(see \refc{TH95_99}, seventh paper)
\begin{eqnarray}
\label{ldgen}
 1 & = &  -  4 i  g^2  N_c \, \int d\!\!{}^{- 4} p
 \; \Big[  p^2 -  \hat{m}^2 \Big]^{-2} \; , \\
1 & = & -  4 i  g^2  N_c \, \int d\!\!{}^{- 4} p \; \Big[\Big(p^2  -
\hat{m}^2 \Big) \; \Big(p^2 -  m^2_s \Big)\Big]^{-1} \; . 
\label{ldges}
\end{eqnarray}
To proceed, given Eqs.~(\ref{fpilsm}) and (\ref{fklsm}), the meson
 charge radii are computed in the L$\sigma$M as 
\begin{eqnarray}
\av{r_{\pi^+}^2}^{\cls}_{\lsms} \; = \; 6 \left.
 \frac{dF_\pi (k^2)}{dk^2} \right|_{k^2 = 0} & = & \frac{ - i 4 N_c 
 g^2 \, (- 2)}{(2\pi)^4} \,\int_0^1 dx \, 6  x  (1-x) \,\int d\!\!{}^{- 4}
 p \,\, \Big[ p^2 - \hat{m}^2  \Big]^{-3} \nonumber \\
 & = & \frac{ 8  i   N_c}{(2\pi)^4} \, g^2 \, (\frac{-i\pi^2}{2
 \hat{m}^2}) \; = \; \frac{N_c}{4\pi^2 f^2_\pi} \; \approx \; (0.61\:
 \mbox{fm})^2
\label{rpilsm}
\end{eqnarray}
and (the obvious $SU(3)$ extension)
\begin{eqnarray}
\av{r_{K^+}^2}^{\cls}_{\lsms} \; =  \; 6 \left.
 \frac{dF_K (k^2)}{dk^2} \right|_{k^2 = 0} & = & \frac{ - i 4 N_c  g^2
 \, (- 2)}{(2\pi)^4} \,\int_0^1 dx \, 6 x  (1-x) \int d\!\!{}^{- 4} p \,
 \Big[ p^2 - m_{us}^2 \Big]^{-3} \nonumber \\
 & = & \frac{ 8  i   N_c}{(2\pi)^4} \, g^2 \,
 (\frac{-i\pi^2}{2 m_{us}^2}) \; = \; \frac{N_c}{4\pi^2 f^2_K} \;
 \approx \; (0.49\: \mbox{fm})^2 \, .
\label{rklsm}
\end{eqnarray}
Here we have evaluated the charge radii in the CL \cite{DS95,CS81,BRS98}, with
 $f^{\cls}_\pi \approx 90$ MeV, $f^{\cls}_K \approx 110$ MeV. 

At this point we may return to the perturbative QT results (\ref{fpiqt}) and
(\ref{fkqt}), from which we derive in the CL
\begin{eqnarray}
\label{rpiqt}
\av{r_{\pi^+}^2}^{\cls}_{\qts} & = & \frac{g^2 
 N_c}{4 \pi^2 \hat{m}^2} \quad \stackrel{!}{=} \quad
\av{r_{\pi^+}^2}^{\cls}_{\lsms} \; , \\
 & & \nonumber \\
\av{r_{K^+}^2}^{\cls}_{\qts} & = & \frac{g^2 
 N_c}{4 \pi^2 m_{us}^2} \nonumber \\
 & & \nonumber \\
 & & \times \; \frac{1}{12} \; \frac{1}{( m^2_s - \hat{m}^2)^2} \;  
 \Bigg\{ \hat{m}^4 -  3 \hat{m}^3  m_s - 5 \hat{m}^2   m^2_s  -
  15 \hat{m}  m_s^3 - 2  m^4_s \nonumber \\
 & & \nonumber \\
 & & \qquad\qquad\qquad\qquad
  + \; 2\,\frac{\hat{m}^6 + 3\hat{m}^5 m_s + 6 \hat{m}  m_s^5 +
  2 m_s^6}{m^2_s - \hat{m}^2}  \; \ln\left| \,  \frac{m^2_s}{\hat{m}^2} \,
 \right|  \Bigg\} \nonumber \\ & & \nonumber \\
 & = & \frac{g^2  N_c}{4 \pi^2 \hat{m}^2} \; \left( 1  -
  \frac{5}{6}  \delta + \frac{3}{5}  \delta^2 -  \frac{4}{9} 
 \delta^3 + \frac{22}{63}  \delta^4 -  \frac{2}{7}  \delta^5 + \ldots 
 \right) \; ,
\label{rkqt}
\end{eqnarray} 
with $m_s = (1+ \delta)  \hat{m}$, i.e., $\delta = (m_s/\hat{m}) -  1
\approx 0.44$. The coefficients of the presented Taylor expansion in the
SU(3)-breaking parameter $\delta$ coincide with the ones given in \refc{AB87},
while our full result is also in agreement with the expressions originally
derived  by Tarrach \cite{T79}. Taking into account the first three terms of
this expansion, we may estimate the ratio $r_K/r_\pi$ to be
\begin{equation} \frac{\av{r_{K^+}^2}}{\av{r_{\pi^+}^2}} \;\approx\; 1 \, - \,
\frac{5}{6} \, \delta + \frac{3}{5} \, \delta^2 \; \approx \; 0.750 \quad
\mbox{or} \quad \frac{\av{r_{K^+}}}{\av{r_{\pi^+}}} \; \approx \; 0.866 \; .
\label{rkdrpi}
\end{equation}
Here we note that the observed pion charge radius is \cite{D82}
\begin{equation}
r_\pi \; = \; (0.642 \pm 0.002)\ \mbox{fm}\; ,
\label{rpiex}
\end{equation}
and the analogue charged-kaon charge radius is \cite{PDG02}
\begin{equation}
r_K \; = \; (0.560 \pm 0.031)\: \mbox{fm}\; .
\label{rkex}
\end{equation}
If we take the experimental value $r_{\pi^+}\approx0.64$ fm from \eqr{rpiex},
the latter ratio (\ref{rkdrpi}) implies $<r_{K^+}> \; \approx \; 0.556$ fm,
which is compatible with \eqr{rkex}.

In summary, the more detailed perturbative results of Eqs.~(\ref{fpiqt}),
(\ref{fkqt}), (\ref{ikmqq}), (\ref{fpiqt0}), (\ref{fkqt0}), (\ref{rpiqt}),
and (\ref{rkqt}) are compatible with the simpler non-perturbative
(SU(3)-symmetry) scheme of Eqs.~(\ref{fpilsm})--(\ref{fklsm0}), (\ref{rpilsm}),
and (\ref{rklsm}) above. Thus, no
further renormalization needs be considered in either case. Note, too, that
these detailed or simple field-theory versions of the charged-pion form factor
can be recovered in an even simpler fashion by using a once-subtracted
dispersion relation for the pion charge radius, yielding in the CL
\begin{equation}
 r^2_\pi \; = \; \frac{6}{\pi} \, \int_0^\infty \frac{dq^2 \,
\Im F_\pi(q^2)}{(q^2)^2} \; = \; \frac{N_c}{4\pi^2
(f^{\cls}_\pi)^2} \; = \; \frac{1}{\hat{m}^2} \; ,
\label{rpicl}
\end{equation}
where we use \cite{DS95} the GTRs \eqr{gtrs}, along with $g=2\pi/\sqrt{N_c}$
from \eqr{comp}. This suggests that the tightly bound ``fused'' $\bar{q}q$ pion
charge radius in the CL is
\begin{equation}
r^{\cls}_\pi \; = \; \frac{1}{\hat{m}} \; = \;
\frac{197.3 \; \mbox{MeV fm}}{325 \; \mbox{MeV}} \; \approx \; 0.61\:
\mbox{fm} \; ,
\end{equation}
with $\hat{m}_{\cls} \approx 325$ MeV $\sim M_N/3$, as
expected from the GTR $m_{\cls}  =  f^{\cls}_\pi
g \approx 90$ MeV $\times 3.628 \; \approx \; 325$ MeV. 
\section{Charged-pion polarizabilities for \bm{\gamma\gamma\rightarrow\pi\pi}}
For $\gamma\gamma\rightarrow\pi\pi$ low-energy scattering, and using
units $10^{-42}$ cm$^{3}$ and effective potential $V=-(\alpha_{\pi}E^2+
\beta_{\pi}B^2)/2$, Kaloshin \em et al.\ \em \/extracted the observed charged
(c) electric and magnetic polarizabilities as
\bea
(\alpha-\beta)^c & = & 6.6 \pm 1.2 \; \cite{KS94} 
\label{amb} \\
(\alpha+\beta)^c & = & 0.37 \pm 0.08 \; \cite{KPS94,KPS95} \; ,
 \; 0.23 \pm 0.09 \; \cite{KPS94,KPS95} \; ,
\label{apb}
\eea
i.e., $\alpha\!-\!\beta$ by employing a combined fit to Crystal-Ball
\cite{CRYSTAL90} and MARK-II \cite{MARKII90} data, and $\alpha\!+\!\beta$ by
fitting CELLO \cite{CELLO92} and MARK-II data, respectively. Adding
\eqrs{amb}{apb} gives
\be
\alpha^c \; = \; 3.45 \pm 0.60 \; .
\label{ac}
\ee
To compare this ``form factor'' to theoretical form-factor predictions, we
first use $\alpha=e^2/4\pi$ and scale up the potential by $4\pi$. Then
$\alpha^c$ in \eqr{ac} becomes
\be
\alpha_{\pi^+} \; = \; (2.75 \pm 0.50) \times 10^{-4}\:\mbox{fm}^3 \;.
\label{apiex}
\ee
Using the latter scale, the model-independent value is \cite{T73}
\be
\alpha_{\pi^+} \; = \; \frac{\alpha}{8\pi^2m_{\pi}f^2_{\pi}}\:\gamma \; ,
\label{apith}
\ee
where $\gamma\equiv F_A(0)/F_V(0)$ is a form-factor ratio found in Sec.~5 to
be $2/3$ in the \lsm. Thus,
\be
\alpha_{\pi^+}^{\lsms} \; = \; \frac{\alpha}{12\pi^2m_{\pi}f^2_{\pi}}
\; \approx \; 3.9 \times 10^{-4}\:\mbox{fm}^3
\label{apilsm}
\ee
is reasonably near the data in \eqr{apiex} above. It is, moreover, quite close
to e.g.\ the prediction $3.6 \times 10^{-4}\:\mbox{fm}^3$ of a quark
confinement model that also yields good results for heavy-meson semi\-leptonic
form factors \cite{IM92}.  Another consistency check is the
detailed quark-plus-meson-loop analysis of \refc{L81}:
\be
\alpha_{\pi^+}^{\lsms} \; = \; \frac{\alpha}{8\pi^2m_{\pi}f^2_{\pi}} \; - \;
\frac{\alpha}{24\pi^2m_{\pi}f^2_{\pi}} \; = \;
\frac{\alpha}{12\pi^2m_{\pi}f^2_{\pi}} \; ,
\label{apilsmd}
\ee
requiring $\gamma^{\lsms}=2/3$ from \eqr{apith}.

Finally we comment on low-energy $\gamma\gamma\rightarrow2\pi^0$ scattering,
where there is no pole term, and the neutral polarizabilities $\alpha_{\pi^0}$,
$\beta_{\pi^0}$ are much smaller than $\alpha_{\pi^+}$, $\beta_{\pi^+}$. In
\refc{KS86} it was shown that a $\gamma\gamma\rightarrow2\pi^0$ cross section
of $\sim10$ nb (generated by a $\sigma$(700) meson pole) reasonably anticipated
the later 1990 Crystal-Ball data \cite{CRYSTAL90} in the 0.3--0.7 GeV range.

\section{Semileptonic weak \bm{K_{\ell3}} decay and form-factor scale
\bm{f_+(0)}}
The semileptonic weak  $\slw$ ($K_{\ell3}$) decay width is measured as
\cite{PDG02}
\be
\Gamma(\slw) \; = \; \frac{\hbar}{\tau_{K^+}}\,(4.87\pm0.06)\% \; = \;
(25.88\pm0.32)\times10^{-16}\:\mbox{MeV} \;.
\label{slwex}
\ee
Taking a $q^2$ form-factor dependence $f_+(q^2)=
f_+(0)[1+\lambda_+q^2/m^2_{\pi}]$, the standard $V\!-\!A$ (vector here) weak
current predicts a $K_{\ell3}$ decay width ($y = m^2_{\pi^0}/m^2_{K^+}$,
$m_e = m_{\nu} = 0$; see also Ref.~\cite{S91})
\begin{eqnarray}  \lefteqn{\Gamma(K^+ \rightarrow \,\pi^0 e^+ \, \nu) \; =
 \; \frac{G_F^2 \, |V_{us}|^2 \, m^5_{K^+}}{2 \, \pi^3 \; 768} \; f^2_{+} (0)
 \,  \Bigg\{ 1- \, 8 \, y + 8 \, y^3 - \, y^4 - \, 12 \, y^2 \,
 \ln y} \nonumber \\
 & & + \, \Bigg( \frac{2}{5}\, \left( 1 - \, 15\, y - \, 80 \, y^2 +
 \, 80\, y^3 +  15\, y^4 - \, y^5 \, \right)  \, -\, 24 \, y^2 \,
 \left( 1 + \, y \, \right)\, \ln y  \Bigg) 
 \; y^{-1} \; \lambda_{+} \nonumber \\
 & &  +  \Bigg( \frac{1}{15}\, \left( 1 - \, 24\, y - \, 375 \, y^2 +
 \, 375\, y^4 +  24 \, y^5 - \, y^6 \, \right)  -\, 4 \, y^2 \,
 \left( 3 + \, 8\,y + 3\, y^2 \, \right)\, \ln y  \, \Bigg)
 \; y^{-2} \;  \lambda^2_{+} \Bigg\} \nonumber \\
 & = & \frac{G_F^2\,|V_{us}|^2\, m^5_{K^+}}{2 \, \pi^3 \, 768} \,
 f^2_{+} (0) \Big( 0.5792  \,+ \; 0.1600 \;
 \frac{m_{K^+}^2}{m^2_{\pi^0}} \; \lambda_{+} \,  + \; 0.01770 \;
 \frac{m_{K^+}^4}{m^4_{\pi^0}}  \;  \lambda^2_{+} \Big)
\nonumber \\ & = & f^2_{+} (0) \;
 (25.90 \pm 0.07)\times 10^{-16}\;\mbox{MeV}  \; ,
\label{slwth}
\end{eqnarray}
where $G_F=11.6639\times10^{-6}$ GeV$^{-2}$, $V_{us}=0.2196\pm0.0026$,
and $\lambda_+= 0.0278\pm0.0019$ \cite{PDG02}. If we neglect here the term
quadratic in $\lambda_+$, as e.g.\ done in Ref.~\cite{S91}, the leading
factor in \eqr{slwth} becomes 25.80 instead of 25.90.
Moreover, accounting for a nonvanishing electron mass yields a totally
negligible correction of the order of 0.001\%. In any case, comparison with
the data in \eqr{slwex} clearly shows that the form-factor scale $f_+(0)$ must
be near unity. However, electroweak radiative corrections to $\Gamma(\slw)$
are \em not \em \/negligible on the scale of the experimental errors in
$V_{us}$ and $\lambda_+$, giving rise to an enhancement of $|V_{us}|$ by more
than 2\% \cite{MS93}, suggesting that $f_+(0)$ should be a trifle less
than unity.

As a matter of fact, the nonrenormalization theorem \cite{AG64} \em requires
\em \/the form factor $f_+(q^2)$ to be close to unity when $q^2=0$.
Furthermore, in the infinite-momentum frame (IMF), tadpole graphs are
suppressed and so
\cite{FF65}
\be
1 - f^2_+(0) \; = \; {\cal O}(\delta^2) \; \approx \; 6\%
\label{fpmo}
\ee
is second order in $SU(3)$-symmetry breaking. Of similar order are, for
example, $(m_\pi/m_K)^2=7.7\%$,
and $(1-f_K/f_\pi)^2=5\%$, for $f_K/f_{\pi}=1.22$.

Next we follow the (constituent) quark-model triangle graph of Fig.~2, with
\begin{figure}[ht]
\unitlength1cm
\epsfxsize=  5.5cm
\epsfysize=  4cm
\centerline{\epsffile{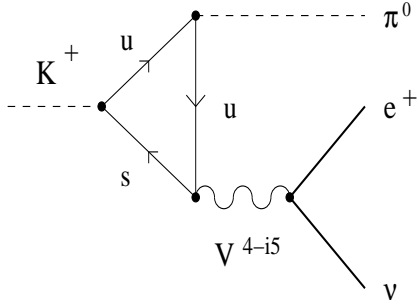}}
\caption{Quark-loop contribution to $K^+ \rightarrow \pi^0 \,\, e^+ \,\nu$.}
\label{figkpienu1}
\end{figure}
\be
\sqrt{2}\bko{\pi^0}{V_{\mu}^{4-i5}}{K^+} \; = \; f_+(t)(p_K+p_{\pi})_{\mu} \, +
\, f_-(t)(p_K-p_{\pi})_{\mu} \; .
\label{vfif}
\ee
Note that, for this process, the $f_-$ form factor can be disposed of, since
it is weighted by $m_e\ll m_K$ \cite{S91}, giving rise to a $m^2_e/m^2_K$
suppression of the corresponding contributions to $\Gamma(\slw)$.
To test $SU(2)$-symmetry breaking in $K_{\ell3}$ decays as in
\eqrs{slwth}{vfif} above, we note the present data consistency \cite{PDG02} of
$\lambda_+(K^+_{e3})=0.0278\pm0.0019$, $\lambda_+(K^0_{e3})=0.0291\pm0.0018$,
$\lambda_+(K^+_{\mu3})=0.033\pm0.010$, and $\lambda_+(K^0_{\mu3})=
0.033\pm0.005$. Then, expanding in the $SU(3)$-breaking parameter $\delta=
(m_s/\hat{m})-1$ (as already used to obtain \eqr{rkdrpi}) and working in the
soft-pion CL, the Feynman graph of Fig.~\ref{figkpienu1} predicts \cite{PS84}
(recall the value of the meson-quark coupling $g\approx3.628$ in \eqr{comp})
\be
f_+(0) \; = \; 1 \, - \, \frac{g^2\delta^2}{8\pi^2} \; \approx \;
0.968 \; .
\label{fkp0}
\ee
This value slightly below unity is not only in agreement with the
nonrenormalization theorem \eqr{fpmo}, as $1-f^2_+(0)=1-(0.968)^2=6.3\%$, but
also quantitatively compatible with \eqrs{slwex}{slwth}, if we account for the
mentioned radiative corrections contributing with about $-2\%$ to $f_+(0)$, and
the experimental errors in $V_{us}$ and $\lambda_+$.

\section{Semileptonic weak radiative form factors for \bm{\pi^+\rightarrow
e^+\nu\gamma} and \bm{K^+\rightarrow e^+\nu\gamma}}

From \refc{PDG02}, the $\pi^+\rightarrow e^+\nu\gamma$ and $K^+\rightarrow
e^+\nu\gamma$ matrix elements are
\bea
M_V & = & \frac{-e\,G_F\,V_{qq'}}{\sqrt{2}\,m_P} \, \epsilon^\mu\ell^\nu
F^P_V\, \epsilon_{\mu\nu\sigma\tau}\,k^\sigma q^\tau \; ,
\label{mv} \\[2mm]
M_A & = & \frac{-ie\,G_F\,V_{qq'}}{\sqrt{2}\,m_P} \, \epsilon^\mu\ell^\nu
\{F^P_A\, [(s-t)g_{\mu\nu}-q_\mu k_\nu] + R^P t\,g_{\mu\nu} \} \; ,
\label{ma}
\eea
where $V_{qq'}$ is the corresponding Cabibbo-Kobayashi-Maskawa (CKM)
mixing-matrix
element, $\epsilon^\mu$ is the photon polarization vector, $\ell^\nu$ is the
lepton-neutrino current, $q$ and $k$ are the meson and photon four-momenta,
respectively, with $s=q\cdot k$, $t=k^2$, and $P$ stands for $\pi$ or $K$.
The weak vector (pion) form factor $F^\pi_V$ in \eqr{mv} and the second
axial vector form factor $R^\pi$ in \eqr{ma} are model independent
\cite{VI58}, with $F^\pi_V$ determined only by conserved vector
currents (CVC), and $R^\pi$ related via the pion charge radius ($r_{\pi}=
0.642\pm0.002$ fm) to partially conserved (pion) axial currents (PCAC).
Specifically, $F^\pi_V$ was long ago determined by CVC \cite{VI58},
viz.\
\be
F^\pi_V(0) \; = \; \frac{\sqrt{2}m_{\pi^+}}{8\pi^2f_{\pi}} \; \approx \;
0.027 \; ,
\label{fpiv0}
\ee
reasonably close to data \cite{PDG02} $0.017\pm0.008$. Furthermore, PCAC
predicts (PCAC is manifest in the \lsm\ \cite{GML60,AFFR73})
\be
R^\pi \; = \; \frac{1}{3}m_{\pi^+}f_{\pi^+}r^2_{\pi^+} \; = \; 0.064\pm0.001\;,
\label{Rpi}
\ee
where $f_{\pi^+}=130.7\pm0.1$ MeV \cite{PDG02} and we use
$r_{\pi^+}=0.642\pm0.002$ fm. Then \eqr{Rpi} is near data \cite{E89}
$R^\pi=0.059\raisebox{-0.7mm}{$\stackrel{+0.009}{\scriptstyle -0.008}$}\:$.

\begin{figure}[ht]
\unitlength1cm
\epsfxsize=  11.5cm
\epsfysize=  4cm
\centerline{\epsffile{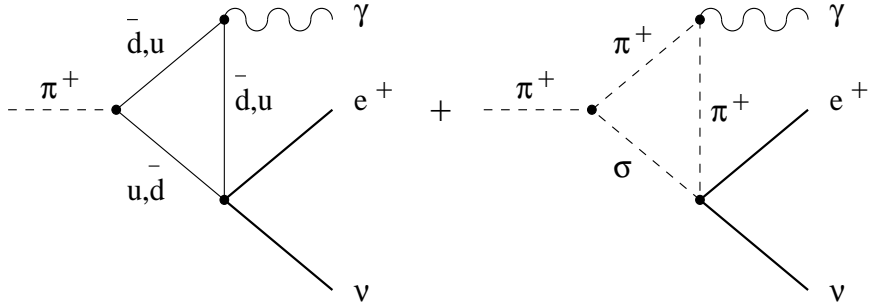}}
\caption{Quark- and meson-loop contribution to
$\pi^+\rightarrow\gamma\,\,e^+\,\nu$.}
\label{figpigamenu1}
\end{figure}

To apply the \lsm\ theory, we consider the quark-plus-meson-loop graphs of
Fig.~\ref{figpigamenu1}.
Then the ratio $\gamma=F_A(0)/F_V(0)$ is predicted as \cite{BS92}
\be
\gamma^{\lsms} \; = \; 1 - \frac{1}{3} \; = \; \frac{2}{3} \; , \\[-5mm]
\label{glsm}
\ee
with
\be
F^\pi_A(0) \; = \; \sqrt{2}\,m_\pi[(8\pi^2f_\pi)^{-1}-(24\pi^2f_\pi)^{-1}] \; =
\; \sqrt{2}\,\frac{m_\pi}{12\pi^2f_\pi} \; \approx \; 0.0179 \; .
\label{fpia0}
\ee
Thus, the form-factor ratio of \eqr{fpia0} divided by \eqr{fpiv0} gives
$\gamma^{\lsms}=0.0179/0.027\approx0.66$, compatible with \eqr{glsm} and with
data \cite{PDG02}:
\be
\gamma_{\mbox{\scriptsize data}} \; = \; \frac{0.0116\pm0.0016}{0.017\pm0.008}
 \;=\; 0.68 \pm 0.33 \; .
\label{gdata}
\ee
With hindsight, this ratio $\gamma^{\lsms}=2/3\,$ is near the original
current-algebra (CA) estimate $0.6$ found in \refc{DMO67}, and exactly the same
$\gamma$ found in \eqr{apith} from the \lsm\ \eqr{apilsmd}.

Extending the above \lsm\ picture to $SU(3)$ symmetry, we first assume a
scalar nonet pattern below 1 GeV (e.g.\ $f_0$(600), $\kappa$(800),
$f_0$(980), $a_0$(980)) as found from a kinematic IMF scheme \cite{S82_92},
or from a dynamical coupled-channel unitarized model \cite{BR86}. Then the
$K^+\rightarrow e^+\nu\gamma$ quark-plus-meson \lsm\ form-factor loop
amplitudes predict \cite{KSB93} at $k^2=0$
\be
|F^K_V(0)+F^K_A(0)|_{\lsms} \; \approx \; 0.109+0.044 \; = \; 0.153 \; ,
\label{fkva0lsm}
\ee
close to the $K^+\rightarrow e^+\nu\gamma$ data \cite{PDG02}
\be
|F^K_V(0)+F^K_A(0)|_{\mbox{\scriptsize data}} \; = \; 0.148\pm0.010 \;.
\label{fkva0exp}
\ee
An $SU(3)$ \lsm\ theory is reasonably detailed \cite{DS98} due to resonances
below 1 GeV, but the \lsm\ kaon form-factor sum in \eqr{fkva0lsm} is easily
tested via the data in \eqr{fkva0exp}. The same is true for the pion
form-factor values in Eqs.~(\ref{fpiv0}--\ref{gdata}), partly based on the
measured pion charge radius \cite{D82} $r_\pi=0.642\pm0.002$ fm.

\section{Vector-meson dominance: \lsm\ via VPP and VPV or PVV loops}
We first confirm the (crucial) value of the pion charge radius \cite{D82}
$r_\pi=0.642\pm0.002$ fm via Sakurai's vector-meson-dominance (VMD) prediction
\cite{S60}
\be
r_\pi \; = \; \frac{\sqrt{6}}{m_\rho} \; \approx \; 0.63\:\mbox{fm} \; .
\label{rpivmd}
\ee
Recall that the tightly bound $\bar{q}q$ chiral pion in \eqr{rpicl}, with
constituent
quark mass $\hat{m}\approx325$ MeV (near $\hat{m}\approx M_N/3$), has CL charge
radius $r^{\cls}_\pi=1/\hat{m}\approx0.61$ fm. So the close agreement between
\eqrs{rpivmd}{rpicl} means we must take the VMD scheme along with the \lsm\ as
the basis of our chiral theory.

\begin{figure}[ht]
\unitlength1cm
\epsfxsize=  11.5cm
\epsfysize=  4cm
\centerline{\epsffile{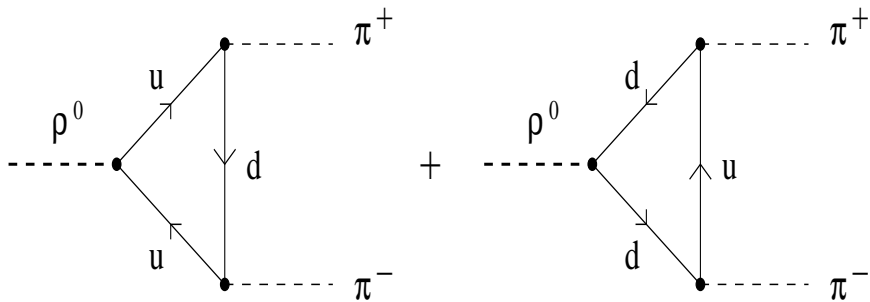}}
\caption{Vector-mesonic VPP quark triangle graphs.} \label{figrhopipi1}
\end{figure}

\begin{figure}[ht]
\unitlength1cm
\epsfxsize=  15.0cm
\epsfysize=  3.5cm
\centerline{\epsffile{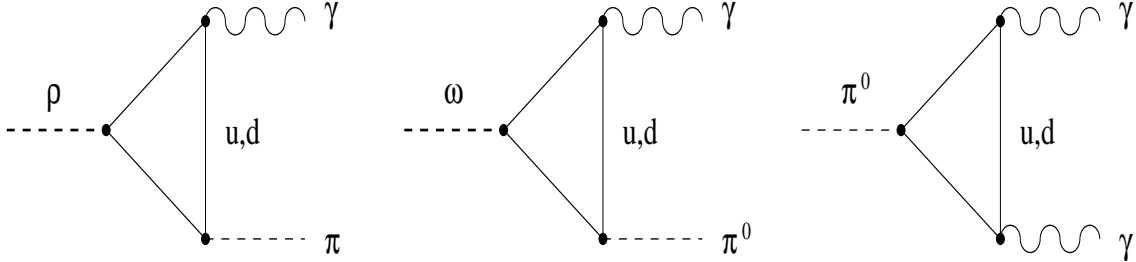}}
\caption{PVV quark triangle graphs for $\rho\rightarrow \pi\,\gamma$,
$\omega\rightarrow\gamma\,\pi^0$, and $\pi^0\rightarrow \gamma\,\gamma$.}
\label{figvpigam1}
\end{figure}

The $\rho^0$ form factor predicts, from $udu+dud$ quarks loops in the CL
(see Fig.~\ref{figrhopipi1}),
\be
g_{\rho\pi\pi} \; = \; -i4N_c\,g^2g_\rho\,\int\dbarfp\,(p^2-\hat{m}^2)^{-2} \;
= \; g_\rho \; ,
\label{grhopipi}
\ee
by virtue of the LDGE \eqr{ldgen} \cite{HS91}. Then, folding in the mesonic
$\pi$-$\sigma$-$\pi$ loop changes the VMD prediction (\ref{grhopipi}) only 
slightly to \cite{BRS98}
\be
g_{\rho\pi\pi} \; = \; g_\rho + \frac{1}{6}\:g_{\rho\pi\pi} \; = \;
\frac{6}{5}\: g_\rho \; ,
\label{grho}
\ee
compatible with the observed couplings $g_{\rho\pi\pi}\approx6.04$ and
$g_\rho\approx5.01$, since (for $p_{\cms}=358$ MeV)
\bea
\Gamma_{\rho\pi\pi} & = & \frac{p^3_{\cms}\,g^2_{\rho\pi\pi}}{6\pi m^2_{\rho}}
\; = \; 149.2\pm0.7 \:\mbox{MeV} \;\; \Longrightarrow \;\; g_{\rho\pi\pi}
\approx 6.04 \\
\Gamma_{\rho ee}    & = & \frac{e^4\,m_\rho}{12\pi g^2_\rho} 
\; = \; 6.85\pm0.11 \:\mbox{keV} \;\; \Longrightarrow \;\; g_\rho \approx
5.01 \; ,
\label{Grhopie}
\eea
with $e\approx0.3028$ (i.e., $\alpha\approx1/137$). Also, the quark-loop VPV or
PVV (see Fig.~\ref{figvpigam1}) amplitudes are \cite{DLS99}, using
$\Gamma_{VPV}=p^3|F_{VPV}|^2/12\pi$,
\be
\begin{array}{l} \displaystyle 
|F(\rho\!\rightarrow\!\pi\gamma)| \; = \; \frac{eg_\rho}{8\pi^2f_\pi}
\; \approx \; 0.207\:\mbox{GeV}^{-1} \;\;,\;\;
|F(\omega\!\rightarrow\!\pi\gamma)| \;  = \; \frac{eg_\omega}{8\pi^2f_\pi}
\; \approx \; 0.704\:\mbox{GeV}^{-1} \;\;,\;\; \\[2mm] \displaystyle 
|F(\pi^0\!\rightarrow\!2\gamma)| \; = \; \frac{\alpha}{\pi f_\pi} \; = \;
\frac{e^2}{4\pi^2f_\pi} \; \approx \; 0.025\:\mbox{GeV}^{-1} \; ,
\label{frhoomegapi}
\end{array}
\ee
for $g_\rho\approx5.01$ and $g_\omega\approx17.06$, very close to the data
$0.222\pm0.012$ GeV$^{-1}$ \cite{PDG02}, $0.698\pm0.014$ GeV$^{-1}$
\cite{BEI96}, $0.0252\pm0.0009$ GeV$^{-1}$ \cite{PDG02},
respectively. Equivalently, VMD predicts at tree level $|F_{\rho\pi\gamma}|\,
e/g_\rho=|F_{\omega\pi\gamma}|\,e/g_\omega=|F_{\pi^0\gamma\gamma}|/2$, then
compatible with the \lsm\ quark loops in \eqr{frhoomegapi}.


\section{Nonleptonic weak \bm{K_{2\pi}$ $\Delta I\!=\!1/2\,} rule and 
scalar \bm{\sigma}, \bm{\kappa} mesons}
\begin{figure}[ht]
\unitlength1cm
\epsfxsize=  5.5cm
\epsfysize=  3.5cm
\centerline{\epsffile{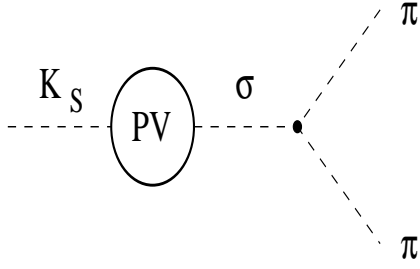}}
\caption{Parity-violating two-pion decay of $K_S$ dominated by $\sigma$ pole.}
\label{figkssigma1}
\end{figure}

The well-known \cite{PDG02} $\Delta I=1/2$ rule
$\Gamma(K_S\rightarrow\pi^+\pi^-)/\Gamma(K^+\rightarrow\pi^+\pi^0)\approx450$
for nonleptonic weak $K_{2\pi}$ decays suggests \cite{MLS90} that the
parity-violating (PV) amplitude $\bko{2\pi}{H^{pv}_w}{K_S}$ could be dominated
by the $\Delta I=1/2$ weak transition $\bko{\sigma}{H^{pv}_w}{K_S}$. The
$\sigma$-pole graph of Fig.~\ref{figkssigma1},
with \lsm\ coupling $\bk{2\pi}{\sigma}=m^2_\sigma/2f_\pi$ for $m_\sigma$ near
$m_K$ and $\Gamma_\sigma\sim m_\sigma$, predicts \cite{KS91}
\be
|\bko{2\pi}{H^{pv}_w}{K_S}| \; = \; \left|\frac{2\bk{2\pi}{\sigma}
\bko{\sigma}{H^{pv}_w}{K_S}}{m^2_K-m^2_\sigma+im_\sigma\Gamma_\sigma}\right| \;
\approx \; \frac{1}{f_\pi}\,\left|\bko{\sigma}{H^{pv}_w}{K_S}\right| \; .
\label{ksigma}
\ee
But pion PCAC (manifest in the \lsm) requires, using the weak chiral commutator
$[Q_5+Q,H_w]=0$,
\be
|\bko{2\pi}{H^{pv}_w}{K_S}|\rightarrow\frac{1}{f_\pi}\,
|\bko{\pi}{[Q_5^\pi,H_w]}{K_S}| \; \approx \;
\frac{1}{f_\pi}\,\left|\bko{\pi^0}{H^{pc}_w}{K_L}\right| \; ,
\label{kskl}
\ee
with both pions being consistently reduced in \refc{KS92}. To reconfirm
\eqr{kskl}, one considers the $\Delta I=1/2$ weak tadpole graph, giving
\be
|\bko{2\pi}{H^{pv}_w}{K_S}| \;=\; \frac{|\bko{0}{H_w}{K_S}
\bk{K_S2\pi}{K_S}|}{m^2_K}\;,
\label{tadpole}
\ee
and then one invokes the Weinberg-Osborn \cite{WO66_70} strong chiral coupling
$|\bk{K_S2\pi}{K_S}|=m^2_{K_S}/2f^2_\pi$, together with the usual PCAC relation
$|\bko{0}{H^{pv}_w}{K_S}|=|2f_\pi\bko{\pi^0}{H^{pc}_w}{K_L}|$, to recover
\eqr{kskl} \cite{PTSE02}.

In either case, equating \eqr{kskl} to \eqr{ksigma} leads to
\be
\left|\bko{\sigma}{H^{pv}_w}{K_S}\right| \; \approx \;
\left|\bko{\pi^0}{H^{pc}_w}{K_L}\right|\;,
\label{ksigmapi}
\ee
suggesting that the $\pi$ and $\sigma$ mesons are ``chiral partners'', at least
for nonleptonic weak interactions. But of course, Secs.~1--6 above also show
that the $\pi$ and the $\sigma$ are chiral partners for strong,
e.m., and semileptonic weak interactions, as well.
To compare this chiral-partner $K\rightarrow\pi$ transition with $K_{2\pi}$
data, we return to the PCAC equation (\ref{kskl}) to write, for $f_\pi\approx
93$ MeV,
\be
|\bko{2\pi}{H^{pv}_w}{K_S}| \; \approx \; 
\frac{1}{f_\pi}\,\left|\bko{\pi^0}{H^{pc}_w}{K_L}\right| 
\; \approx \; 38\times10^{-8}\:\mbox{GeV} \; ,
\label{kskld}
\ee
midway between the observed $K_S\rightarrow\pi^+\pi^-$ and
$K_S\rightarrow\pi^0\pi^0$ amplitudes
\be
\left|M^{+-}_{K_S\rightarrow\pi\pi}\right|_{\mbox{\scriptsize PDG}} \; = \;
m_{K_S}\,\left[\frac{8\pi\Gamma^{K_S}_{+-}}{q}\right]^{\frac{1}{2}} \; = \;
(39.1\pm0.1)\times10^{-8}\:\mbox{GeV} \; ,
\label{mkpipippdg}
\ee
\be
\left|M^{00}_{K_S\rightarrow\pi\pi}\right|_{\mbox{\scriptsize PDG}} \; = \;
m_{K_S}\,\left[\frac{16\pi\Gamma^{K_S}_{00}}{q}\right]^{\frac{1}{2}} \; = \;
(37.1\pm0.2)\times10^{-8}\:\mbox{GeV} \; ,
\label{mkpipi0pdg}
\ee
suggesting $|\bko{\pi^0}{H_w^{pc}}{K_L}|\approx3.58\times10^{-8}$ GeV$^2$.
In fact, when one statistically averages \em eleven \em \/first-order weak
data sets for $K_S\rightarrow2\pi$, $K\rightarrow3\pi$,
$K_L\rightarrow2\gamma$, $K_L\rightarrow\mu^+\mu^-$,
$K^+\rightarrow\pi^+e^+e^-$, $K^+\rightarrow\pi^+\mu^+\mu^-$, and
$\Omega^-\rightarrow\Xi^0\pi^-$, one finds \cite{LS02}
\be
\left|\bko{\pi^0}{H_w^{pc}}{K_L}\right| \; = \;
\left|\bko{\pi^+}{H_w^{pc}}{K^+}\right| \; = \;
(3.59\pm0.05)\times10^{-8}\:\mbox{GeV}^2 \; .
\label{hpik}
\ee

\begin{figure}[ht]
\unitlength1cm
\epsfxsize=  5.5cm
\epsfysize=  4cm
\centerline{\epsffile{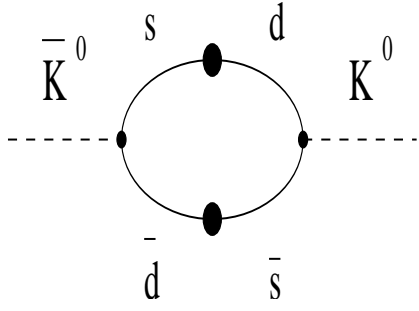}}
\caption{$\bar{K}^0\leftrightarrow K^0$ SQL graph. Each dot represents the SQL
weak scale $\beta_w$.} \label{figkbark1}
\end{figure}

To induce theoretically at the quark level the $\Delta I\!\!=\!\!1/2$
$\:s\rightarrow d$ single-quark-line (SQL) transition scale $\beta_w$ in a
model-independent manner, one considers the second-order weak
(see Fig.~\ref{figkbark1})
$K_L\!-\!K_S$ mass difference $\Delta m_{LS}$ diagonalized to \cite{SEC95_96}
\be
2\beta_w^2 \; = \; \frac{\Delta m_{LS}}{m_K} \; = \;
(0.70126\pm0.00121)\times10^{-14} \;\; \Longrightarrow \;\; |\beta_w| \;
\approx \; (5.9214\pm0.0051)\times10^{-8} \; .
\label{betaw}
\ee
Then using \eqr{hpik}, one predicts from the soft-meson theorem, or from
Cronin's chiral Lagrangian \cite{C67},
\be
\left|\bko{\pi^0}{H^{pc}_w}{K_L}\right|\;=\;2\beta_wm^2_{K_L}\frac{f_K}{f_\pi}
\; = \; (3.5785\pm0.0031)\times10^{-8}\:\mbox{GeV}^2 \; ,
\label{cronin}
\ee
given $f_K/f_\pi\approx1.22$. This SQL scale $\beta_w$ in \eqr{betaw} and the
$K\rightarrow\pi$ weak amplitude in \eqr{cronin} (or in \eqr{hpik}),
correspond to a ``truly weak'' interaction, which Weinberg \cite{W73} shows
cannot be transformed away in the electroweak standard model.

\begin{figure}[ht]
\unitlength1cm
\epsfxsize=  5.5cm
\epsfysize=  3.8cm
\centerline{\epsffile{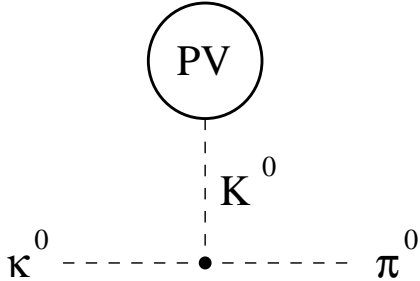}}
\caption{Parity-violating weak $K^0$ tadpole graph for
$\kappa^0\rightarrow\pi^0$ transition.}
\label{kzerotadpole}
\end{figure}

To test the latter weak scale (\ref{cronin}) (or the similar data averages
(\ref{hpik}), we first re-express the neutral chiral-partner relation
(extended to the $\kappa$ transition \cite{KS91}) as
\be
\bko{\pi^0}{H_w^{pc}}{K^0} \: = \: \bko{\sigma}{H_w^{pv}}{K^0} \: = \:
\bko{\pi^0}{H_w^{pv}}{\kappa^0} \: = \:
\frac{1}{\sqrt{2}}\,3.58\times10^{-8}\:\mbox{GeV}^2 \: = \:
2.53\times10^{-8}\:\mbox{GeV}^2 \; .
\label{hpikappa}
\ee
We fix this $\kappa^0\rightarrow\pi^0$ weak transition (\ref{hpikappa}) to
the weak PV $K^0$ tadpole graph of Fig.~\ref{kzerotadpole}, via the
$K^0\rightarrow$ vacuum PCAC scale, as
\be
\left|\bko{0}{H_w^{pv}}{K^0}\right| \; = \; \frac{2f^2_\pi}{1-m^2_\pi/m^2_K}
\,\left|\bko{2\pi^0}{H_w^{pv}}{K^0}\right| \; = \;
0.51\times10^{-8} \: \mbox{GeV}^3 \; ,
\label{ktadpoleexp}
\ee
using $\left|\bko{2\pi^0}{H_w^{pv}}{K^0}\right|=26.26\times10^{-8}$ GeV from
data, while eliminating the 4\% $\,\Delta I\!=\!3/2$ component (see \refc{W73},
third paper). Then Fig.~\ref{kzerotadpole} predicts the amplitude magnitude
\be
\left|\bko{\pi^0}{H_w^{pv}}{\kappa^0}\right| \; = \;
\frac{\left|\bko{0}{H_w^{pv}}{K^0}\right|}{m^2_{K^0}}\:
g_{\kappa^0K^0\pi^0} \; \approx \; 2.53\times10^{-8}\:\mbox{GeV}^2 \; ,
\label{ktadpole}
\ee
scaled to \eqr{hpikappa} above, \em provided \em \/one uses the \lsm\
coupling, for $f_\pi=92.4$ MeV \cite{PDG02},
\be
|g_{\kappa^0K^0\pi^0}| \; = \; \frac{m^2_\kappa-m^2_K}{4f_\pi} \; = \;
1.229 \: \mbox{GeV}\;,
\label{gkappakpi}
\ee
corresponding to a $\kappa$ mass of 838 MeV. This value is not too
distant from our earlier $m_\kappa=$ 730--800 MeV predictions
\cite{S82_92,BR86}, and the very recent E791 observed mass $m_\kappa\approx$
800 MeV \cite{A02}. Moreover, the $SU(3)$ analogue to \eqr{gkappakpi}, i.e.,
$|g_{\sigma\pi\pi}|=(m^2_\sigma-m^2_\pi)/2f_\pi$ suggests $m_\sigma=687$ MeV,
reasonably near the predicted CL-\lsm\ value \cite{DS95,S99} $m_\sigma=650$
MeV.

\section{Summary and conclusions}
In Sec.~1 we reviewed the solution of the \lsm\ at the quark-loop level. In
Sec.~2 we used $SU(2)$, $SU(3)$ Goldberger--Treiman quark relations to
normalize the
$\pi$ and $K$ form factors to unity at $k^2=0$, after which we differentiated
these form factors to predict the \lsm\ charge radii, both being compatible
with data. Next in Sec.~3 we briefly reviewed e.m.\ charged-pion
polarizabilities for $\gamma\gamma\rightarrow\pi\pi$, and compared them with
\lsm\ predictions. In Sec.~4 we used quark loops to match the observed form
factor $f_+(0)$. In Sec.~5 we showed that the \lsm\ form factors $F_V^\pi$,
$R^\pi$, $F_V^K+F_A^K$, and the ratio $F_A^\pi/F_V^\pi$ are all in agreement
with the measured values. Then in Sec.~6 we compared tree-level VMD with \lsm\
VPP and PVV quark loops. Both theories agree well with data. Finally, in Sec.~7
we successfully extended this \lsm\ picture to nonleptonic weak decays, 
in particular to the $\Delta I\!=\!1/2$-dominated $K_{2\pi}$ decays and
inferred $\sigma$(687) and $\kappa$(838) masses. All our main results are
summarized in Table~\ref{results}, in confrontation with experiment.
\begin{table}[ht]
\begin{center}
\begin{tabular}{|l||c|c|}
\hline\hline & & \\ [-0.3cm]
& L$\sigma$M (CL) & experiment \\
\hline & & \\ [-0.3cm]
$\av{r_{\pi^{+}}}$ & 0.61 fm & (0.642$\pm$0.002) fm \cite{D82}
\\ [.25cm]
$\av{r_{K^{+}}}$ & 0.49 fm & (0.560$\pm$0.031) fm \cite{PDG02}
\\ [.25cm]
$\alpha_{\pi^{+}}$ & 3.9$\times 10^{-4}$ fm$^{3}$ &
(2.75$\pm$0.50)$\times 10^{-4}$ fm$^{3}$ \\
 & & (see Sec.~3)\\ [.2cm]
$f_{+}(0)$ & 0.968 & (see discussion in Sec.~4) \\ [.25cm]
$R^{\pi}$ & 0.064$\pm$0.001 &
0.059$^{+0.009}_{-0.008}$ \cite{E89} \\ [.25cm]
$F^{\pi}_{A}$ & 0.0179 & 0.0116$\pm$0.0016 \cite{PDG02} \\ [.25cm]
$F^{\pi}_{V}$ & 0.027 & 0.017$\pm$0.008 \cite{PDG02} \\ [.15cm]
$\gamma^{\pi}\; =\;\fnd{F^{\pi}_{A}(0)}{F^{\pi}_{V}(0)}$ &
$\frac{2}{3}$ & 0.68$\pm$0.33 \cite{PDG02} \\ [.35cm]
$\abs{F^{K}_{V}(0)+F^{K}_{A}(0)}$ & 0.153 &
0.148$\pm$0.010 \cite{PDG02} \\ [.25cm]
$\abs{F\left(\rho\rightarrow\pi\gamma\right)}$ & 0.207 GeV$^{-1}$ &
(0.222$\pm$0.012) GeV$^{-1}$ \cite{PDG02} \\ [.25cm]
$\abs{F\left(\omega\rightarrow\pi\gamma\right)}$ & 0.704 GeV$^{-1}$ &
(0.698$\pm$0.014) GeV$^{-1}$ \cite{BEI96} \\ [.25cm]
$\abs{F\left(\pi^{0}\rightarrow\gamma\gamma\right)}$ & 0.025 GeV$^{-1}$ &
(0.0252$\pm$0.0009) GeV$^{-1}$ \cite{PDG02} \\ [.25cm]
$\abs{\bko{2\pi}{H^{pv}_{w}}{K_{S}}}$ & 38$\times 10^{-8}$ GeV &
$\left.\begin{array}{ll}
(39.1\pm 0.1)\times 10^{-8} \mbox{ GeV} & (\pi^{+}\pi^{-}) \\
(37.1\pm 0.2)\times 10^{-8} \mbox{ GeV} & (\pi^{0}\pi^{0}) \\
\end{array}\right\}$ \cite{PDG02}\\ [.45cm]
$m_{\kappa}$ & 838 MeV & 797$\pm$19$\pm$43 MeV \cite{A02} \\
\hline\hline
\end{tabular}
\end{center}
\caption[]{Confrontation of the \lsm\ in the chiral limit with experiment.}
\label{results}
\end{table}

Next, we discuss low-energy QCD. While an exact match via the \lsm\ is not
possible, QCD at the 1-GeV scale generates a dynamical quark mass
\cite{ESB84_97} $m_{\mbox{\scriptsize dyn}}=[4\pi\alpha_s\left<-\bar{\psi}\psi
\right>_{\mbox{\scriptsize1 GeV}}/3]^{1/3}\approx320$ MeV, near the \lsm\ quark
mass $\hat{m}=2\pi f_\pi/\sqrt{3}\approx325$ MeV in the CL. Such approximate
agreement also holds for the quark condensate as well. Moreover, the frozen
coupling strength in QCD at infrared freeze-out \cite{MS92} is
$\alpha_s=\pi/4$, with $\alpha_s^{\mbox{\scriptsize eff}}=(4/3)\alpha_s=\pi/3$.
This exactly matches the \lsm\ strength $\alpha_{\lsms}=g^2/4\pi=\pi/3$, with
$g=2\pi/\sqrt{3}$. Also, QCD with $\alpha_s(m_\sigma)=\pi/4$ leads to
\cite{ES84} $m^2_\sigma/m^2_{\mbox{\scriptsize dyn}}=\pi/\alpha_s(m_\sigma)
\approx4$, simulating the NJL--\lsm\ value $m^2_\sigma/\hat{m}^2=4\,$ in the CL.
Lastly, the chiral restoration temperature $T_c$ computed in $N_f\!=\!2$
lattice simulations gives \cite{K01} $T_c=173\pm8$ MeV, close to the \lsm\
value \cite{BCSKB85_95} $T_c=2f_\pi\approx180$ MeV in the CL.

To conclude, we mention a very recent large-$N_c$ renormalization-group-flow
analysis of the quark-level \lsm\ \cite{MSPD02}, using the Schwinger
proper-time regularization, which finds (for $f_\pi=93$ MeV) $\lambda=23.6$,
$g=3.44$, $m_q=320$ MeV, and $m_\sigma=650$ MeV, strikingly close to our above
theoretical values $\lambda=8\pi^2/3=26.3$, $g=2\pi/\sqrt{3}\approx3.628$,
$m_q=325$ MeV, and $m_\sigma=650$ MeV, respectively. Therefore, our present
results, as well as our recent findings in \refc{KBRS02}, appear to confirm
the assumption of the authors of \refc{MSPD02}: \em ``We assume the linear 
$\sigma$ model to be a valid description of Nature below scales of 1.5 GeV.''
\em  \\[1cm]
{\bf Acknowledgments.} \\[1mm]
The authors are indebted to A.~E.~Kaloshin for valuable information on pion
polarizabilities.  This work was partly supported by the
{\em Funda\c{c}\~{a}o para a Ci\^{e}ncia e a Tecnologia} (FCT) 
of the {\em Minist\'{e}rio do Ensino Superior, Ci\^{e}ncia e Tecnologia} of 
Portugal, under Grant no.\ PRAXIS XXI/\-BPD/\-20186/\-99 and under contract
number CERN/\-P/\-FIS/\-43697/\-2001.

\clearpage

\appendix
{\noindent \Large\bf APPENDIX}
\section{Tree-level L\bm{\sigma}M}
From the $SU(2)$ \lsm\ of \refc{S01} one knows the interacting Lagrangian
density relative to the true vacuum \cite{DS95}:
\begin{equation}
{\cal L}^{\mbox{\scriptsize int}}_{\mbox{\scriptsize\lsm}} = g\,\bar{\psi}
(\sigma+i\gamma_5\vec{\tau}\cdot\vec{\pi})\psi\,+\,g'\,\sigma\,(\sigma^2+\pi^2)
\,-\,\frac{\lambda}{4}\,(\sigma^2+\pi^2)^2 \, - \, f_\pi g\,\bar{\psi}\psi \; .
\label{lsm}
\end{equation}
A tree-level theory then implies the chiral relations in the CL
\cite{GML60,AFFR73}, with constituent quark mass $m_q$,
\be
g \; = \; \frac{m_q}{f_\pi} \;\;\; , \;\;\; g' \; = \;
\frac{m^2_\sigma}{2f_\pi} \; = \; \lambda\,f_\pi \; .
\label{ggprime}
\ee
\section{Bootstrapping \bm{g_{\sigma\pi\pi}\rightarrow g'} and \bm{
\lambda_ {\mbox{\scriptsize box}}\rightarrow\lambda_{\mbox{\scriptsize tree}}}}
The $\sigma\pi\pi$ or $\sigma\sigma\sigma$ $\,u,d$ \/quark triangle graphs
\cite{DS95,S01} induced by ${\cal L}^{\mbox{\scriptsize int}}_{\lsms}$ in
\eqr{lsm} implies in the CL
\be
g_{\sigma\pi\pi} \; = \; - 8 i g^3  N_c m_q  \int d\!\!{}^{- 4} p\; \Big[ p^2 - 
\hat{m}^2 \Big]^{- 2} \; = \; 2gm_q \; ,  
\label{quarktriangle}
\ee
due to the LDGE (\ref{ldgen}). Then the GTR \eqr{gtrs}, together with
$m_\sigma=2m_q$, reduces \eqr{quarktriangle} to
\be
g_{\sigma\pi\pi} \; = \; 2gm_q \; = \; \frac{m^2_\sigma}{2f_\pi} \; = \; g' \;,
\label{gboot}
\ee
the tree-level cubic meson \lsm\ coupling in \eqr{ggprime}. Also the
$\pi\pi\pi\pi$ (or $\sigma\sigma\sigma\sigma$, $\pi\pi\sigma\sigma$) quark box
graph \cite{DS95,S01} generates in the CL
\be
\lambda_{\mbox{\scriptsize box}} \; = \; - 8 i g^4  N_c  \int d\!\!{}^{- 4}p\;
\Big[ p^2 - \hat{m}^2 \Big]^{- 2} \; = \; 2g^2 \; = \; \frac{g'}{f_\pi} \; = \;
\lambda_{\mbox{\scriptsize tree}} \; ,
\label{quarkbox}
\ee
again due to the LDGE (\ref{ldgen}). Note that the cubic and quartic \lsm\ tree
couplings in \eqr{ggprime} are dynamically loop-generated in
\eqrs{quarktriangle}{quarkbox}, respectively. Both are analytic,
nonperturbative bootstrap procedures \cite{DS95}.

\section{Dim-reg.\ lemma generating quark and \bm{\sigma} mass}
The Nambu $\delta m_q=m_q$ (constituent-) quark mass-gap tadpole graph
\cite{DS95,S01} generates quark mass. However, this quadratically divergent
term, subtracted from the LDGE (\ref{ldgen}), in fact scales to quark mass \em
independently \em \/of quadratically divergent terms, by virtue of the
dimensional-regularization (dim-reg.) lemma \cite{DS95}
\be
I \; = \; \int d\!\!{}^{- 4}p\;\left[\frac{m^2}{(p^2-m^2)^2} \, - \,
\frac{1}{p^2-m^2}\right] \; = \; \lim_{\ell\rightarrow2}\frac{im^{2\ell-2}}
{(4\pi)^\ell}\left[\Gamma(2-\ell)+\Gamma(1-\ell)\right] \; = \;
-im^2(4\pi)^{-2} \; ,
\label{dimreg}
\ee
due to the gamma-function \em identity \em \/$\Gamma(2-\ell)+\Gamma(1-\ell)=
\Gamma(3-\ell)/(1-\ell)\rightarrow-1$ as $\ell\rightarrow2$. To reconfirm this
dim-reg.-lemma ``trick'' (\ref{dimreg}), we invoke the partial-fraction \em
identity \em
\be
\frac{m^2}{(p^2-m^2)^2} \, - \, \frac {1}{p^2-m^2} \; = \; \frac{1}{p^2}\,
\left[\frac{m^4}{(p^2-m^2)^2} \, - \, 1 \right] \; ,
\label{partfrac}
\ee
integrated via $\int d\!\!{}^{- 4}p$ as in the $I$ integral on the l.h.s.\ of 
\eqr{dimreg}. Then dropping the massless-tadpole integral $\int d\!\!{}^{- 4}p/
p^2=0$ (as done in dimensional, analytic, zeta-function, and Pauli--Villars
regularizations \cite{DS95,DSR98}), and Wick rotating
$d^4p=i\pi^2p^2_{\mbox{\scriptsize E}}dp^2_{\mbox{\scriptsize E}}$, the
Euclidean integral becomes
\be
I \; = \; -\frac{im^4}{(4\pi)^2}\int_0^\infty\frac{dp^2_{\mbox{\scriptsize E}}}
{(p^2_{\mbox{\scriptsize E}}+m^2)^2} \; = \; -\frac{im^2}{(4\pi)^2} \; ,
\label{euclidean}
\ee
identical to the r.h.s.\ of \eqr{dimreg}.

In order to further justify the neglect of $\int d\!\!{}^{- 4}p/p^2$, we invoke
the Karlson trick \cite{K54} (long advocated by Schwinger)
\be
\frac{d}{dm^2}\int\frac{d^4p}{p^2-m^2} \; = \; \int\frac{d^4p}{(p^2-m^2)^2}\;,
\label{karlson}
\ee
and compute \cite{DS02}
\be
(2\pi)^4\frac{dI}{dm^2} \; = \; \int\frac{d^4p}{(p^2-m^2)^2} \, + \,
2m^2 \int\frac{d^4p}{(p^2-m^2)^3}\,-\,\frac{d}{dm^2}\int\frac{d^4p}{p^2-m^2}\;,
\label{didm2}
\ee
with the first and third terms cancelling due to \eqr{karlson}. Then the
remaining, finite second term in \eqr{didm2} gives
\be
(2\pi)^4\frac{dI}{dm^2} \; = \; 2m^2\left(-\frac{i\pi^2}{2m^2}\right) \; = \;
-i\pi^2 \; ,
\label{second}
\ee
which is the \em same \em \/result as differentiating the dim-reg.\ lemma
(\ref{dimreg}):
\be
(2\pi)^4\frac{dI}{dm^2} \; = \; (-i\pi^2)\frac{dm^2}{dm^2} \; = \; -i\pi^2 \; .
\label{ddimreg}
\ee
So far we have only assumed $\int d\!\!{}^{- 4}p/p^2$ is independent of $m^2$,
so that $(d/dm^2)\int d\!\!{}^{- 4}p/p^2=0$. 

But to demonstrate that the $\int\!dm^2$ integration constant \em vanishes, \em
i.e., $\int d^4p/p^2=\Lambda^2=0$, we invoke the implied dimensional-analysis
relations
\be
\int\frac{d^4p}{p^2} \; = \; 0 \; , \;\;
\int\frac{d^4p}{p^2-m^2} \; \propto \; m^2 \; , \;\;
\int\frac{d^4p}{p^2-m^2_\sigma} \; \propto \; m^2_\sigma 
\label{dimanal}
\ee
to solve B.~W.~Lee's null-tadpole sum \cite{L72}, which characterizes the true
vacuum for $N_f=2$ as \cite{DS95}
\be
(2m_q)^4N_c \; = \; 3m^4_\sigma
\label{true}
\ee
(with the factor of 3 due to $\sigma$-$\sigma$-$\sigma$ combinatorics)
in the CL $m_\pi=0$, meaning $N_c=3$ when $m_\sigma=2m_q$.

Thus, $\int\!d^4p/p^2$ indeed vanishes as suggested \cite{DS95,DSR98}.
Then appendices A, B, and C loop-generate \eqr{comp} via the LDGE \eqr{ldgen}
and the dim.-reg.\ lemma \eqr{dimreg} \cite{DS95,S01} .

\end{document}